\documentclass[prl, amsfonts, twocolumn, nofootinbib, showpacs]{revtex4} \usepackage{graphicx, epsfig}

\usepackage{amsmath, amssymb}
\usepackage{array}

\begin{document}

\newcommand{\be}{\begin{equation}}
\newcommand{\ee}{\end{equation}}
\newcommand{\pdf}{\mathcal{P}}
\newcommand{\data}{{\bf d}}
\newcommand{\mdl}{\mathcal{M}}
\newcommand{\lsim}{\,\raise 0.4ex\hbox{$<$}\kern -0.8em\lower 0.62ex\hbox{$\sim$}\,} \newcommand{\gsim}{\,\raise 0.4ex\hbox{$>$}\kern -0.7em\lower 0.62ex\hbox{$\sim$}\,}

\newcommand{\params}{\boldsymbol{\theta}}
\newcommand{\paramsU}{\boldsymbol{\theta}_\star}
\newcommand{\fsel}{f_\text{sel}}
\newcommand{\fobs}{f_\text{obs}}
\newcommand{\gal}{\text{gal}}
\newcommand{\stars}{\text{stars}}
\newcommand{\plan}{\text{planets}}
\newcommand{\mpl}{M_\text{Pl}}
\newcommand{\rhol}{\rho_\Lambda}
\newcommand{\rhom}{\rho_m}
\newcommand{\rhor}{\rho_r}
\newcommand{\aeq}{a_\text{eq}}
\newcommand{\Heq}{\mathcal{H}_\text{eq}}
\newcommand{\omleq}{\Omega_{\Lambda, \text{eq}}} \newcommand{\Rmin}{R_\text{min}} \newcommand{\Nmax}{N_\text{max}}

\newcommand{\Tds}{T_\text{dS}}
\newcommand{\mean}{\boldsymbol{\mu}}
\newcommand{\like}{L}
\newcommand{\lnlike}{\mathcal{L}}
\newcommand{\ML}{^*}
\newcommand{\dr}{\textrm{d}}
\newcommand{\ie}{i.e.}
\newcommand{\eg}{e.g.}
\newcommand{\reion}{\text{re}}

\newcommand{\cd}{\cdot}
\newcommand{\cds}{\cdots}
\newcommand{\ip}{\int_0^{2\pi}}
\newcommand{\al}{\alpha}
\newcommand{\ba}{\beta}
\newcommand{\de}{\delta}
\newcommand{\De}{\Delta}
\newcommand{\ep}{\epsilon}
\newcommand{\Ga}{\Gamma}
\newcommand{\ka}{\tau}
\newcommand{\io}{\iota}
\newcommand{\La}{\Lambda}
\newcommand{\Om}{\Omega}
\newcommand{\om}{\omega}
\newcommand{\si}{\sigma}
\newcommand{\Si}{\Sigma}
\newcommand{\te}{\theta}
\newcommand{\ze}{\zeta}
\newcommand{\vth}{\ensuremath{\vartheta}}
\newcommand{\vph}{\ensuremath{\varphi}}
\newcommand{\MM}{\mbox{$\cal M$}}
\newcommand{\tr}{\mbox{tr}}
\newcommand{\hor}{\mbox{hor}}
\newcommand{\grad}{\mbox{grad}}
\newcommand{\cx}{\ensuremath{\mathbf{\nabla}}}

\title[The anthropic principle]{Why anthropic reasoning cannot predict $\Lambda$}

\author{Glenn D. Starkman$^{1,2}$}
\author{Roberto Trotta$^1$}
\affiliation{$^1$ Astrophysics Department, Oxford University,
Denys Wilkinson Building, Keble Road, Oxford OX1 3RH, UK}
\affiliation{$^2$ Department of Physics, Case Western Reserve
University, Cleveland, OH~~44106-7079, USA}

\begin{abstract}
We revisit anthropic arguments purporting to explain the measured
value of the cosmological constant. We argue that different ways
of assigning probabilities to candidate universes lead to totally
different anthropic predictions. As an explicit example, we show
that weighting different universes by the total number of possible
observations leads to an extremely small probability for observing
a value of $\Lambda$ equal to or greater than what we now measure.
We conclude that anthropic reasoning within the framework of
probability as frequency is ill--defined and that in the absence
of a fundamental motivation for selecting one weighting scheme
over another the anthropic principle cannot be used to explain the
value of $\Lambda$, nor, likely, any other physical parameters.

\end{abstract}

\pacs{98.80.-k}

\maketitle

One of the main goals of physics has been to explain the laws of
nature from fundamental principles, such as the existence and
breaking of symmetries. This program has so far been carried out
with great success and attempts are being made to expand it to
explain the values of the dimensionless constants that arise in
our theories. Several of the most outstanding ``problems'' in
fundamental physics of the last decades have been how to explain
particularly deviant dimensionless numbers: the tiny ratio of the
weak energy scale to the Planck mass (``the gauge hierarchy
problem''), the large entropy of the universe (``the entropy
problem''), the tiny value of the geometric curvature of space in
Planck units when evolved back to the Planck time (``the flatness
problem'') and the incredibly low energy density of the vacuum
compared to the characteristic Planck energy density ( ``the
cosmological constant problem'').

As an alternative to explanation from fundamental principles, a
form of probabilistic reasoning known as the ``anthropic
principle'' has become popular, especially as applied to the
cosmological constant problem~\cite{Weinberg:1988cp_etal}. The
argument can be summarized as follows. One assumes that the values
taken on in our Universe by the constants of Nature are just one
realization of a statistical ensemble. This ensemble may be
thought of as causally disconnected patches of this Universe, as
separate sub--universes (a.k.a. the multiverse) or as a
superposition of states (in quantum cosmology). Given some {\it a
priori} distribution of the values of the fundamental constants
across the ensemble, the probability for a ``typical'' observer to
measure a certain value of one or more of these constants is
usually taken to be proportional to the number of observers (or
the number of observers per baryon) that develop in a universe
with that value. Values of the fundamental constants that are
incompatible with the development of intelligent life will never
be observed. In the case of the cosmological constant $\Lambda$,
or equivalently the vacuum energy density $\rho_\Lambda$, this
selection effect is claimed to successfully predict $\rhol$
comparable to what is actually observed, ie $\rhol/\mpl^4 \approx
10^{-123}$ \cite{Weinberg:1988cp_etal}.

As we cannot determine the actual number (density) of observers in
our own Universe, not to mention in some hypothetical universe
with different constants of nature, a more readily calculated
surrogate is used -- the physical number density of galaxies. We
may question whether this is indeed the appropriate weighting
factor. For one thing, since it is a function of time, we must
choose when to evaluate it. For another, we might be concerned
that it fails to differentiate between universes in which a
typical galaxy lasts for a very short time (say one which
recollapses after a billion years) and universes in which a
typical galaxy persists for a very long time (perhaps trillions of
years, or longer).

We shall therefore argue that there are other plausible weighting
factors for universes, and that the answers to questions such as
the expected value of $\Lambda$ depends enormously on the
weighting.As an example, we introduce a weighting scheme based on
the maximal number of allowed observations (MANO) in a universe.
This quantity is clearly relevant to the expected value of a
constant, say $\Lambda$, since a value that allows more
observations to be carried out will be measured more often. It
also has the advantage of being time-independent. In this {\em
Letter}, we show that in this approach the anthropically predicted
probability of measuring $\rhol$ to be greater than or equal to
the currently inferred value, in a universe otherwise similar to
our own, is $\sim10^{-5}$, in marked contrast to the usual result.
The result also depends on other quantities that are effectively
unknowable because they describe complex emergent phenomena.
However, even in the very optimistic case of an early emergence of
intelligent life, the probability of measuring a large $\Lambda$
is still $\sim 5 \cdot 10^{-4}$.

While we do not argue that our probabilistic weighting scheme is
better than the traditional one, it is certainly no worse. Since
the conclusions one derives depend enormously on which weighting
scheme one uses, we conclude that anthropic reasoning based on
such frequency arguments is ill--defined. It cannot be used to
explain the value of $\Lambda$, nor, likely, any other physical
parameters.

We focus on the case where only the cosmological constant varies
from one realization to another. We keep fixed all other
fundamental constants, as well as all remaining cosmological
parameters. This approach has been widely used in the literature.
(See \cite{Aguirre:2001zx} for a discussion of how the situation
changes when more parameters are varied.)

As described above, the value that $\Lambda$ takes on in our
Universe is seen as the outcome of a sampling from a fundamental
probability distribution $f(\Lambda)$. The probability of
observing a specific value $\Lambda$ is then \be \label{eq:prob}
\fobs(\Lambda) = f(\Lambda)\fsel(\Lambda)  . \ee $\fsel(\Lambda)$
is the probability of the observation. It encapsulates the
selection effects, giving different weights to different
universes. There are at least two shortcomings to this approach
(see also \cite{Aguirre:2005cj}).

The first point is that the very concept of probability as a
limiting frequency of outcomes, though natural when applied to
repeatable experiments, is not obviously applicable to the whole
Universe. One solution is to use ergodic arguments to derive $f$
\citep{Tegmark:2005dy}. The validity of this approach remains
unproven. More radical is the multiverse scenario, according to
which there is an infinite collection of, by definition
inaccessible, universes.  This approach hardly seems economical.
It is difficult to see how vastly increasing the number of
universes could help determine the properties of the one Universe
we actually observe. We will not address issues relating to $f$
any further, but will make the usual assumption that $f(\Lambda)$
is flat within the region where $\fsel$ is non--vanishing.

The second aspect is that the selection effect probability $\fsel$
is strongly dependent on the way one chooses to give weight to
different universes. We introduce below a new possibility, namely
the maximum possible number of observations (MANO) over the entire
life of the universe. This concept has the advantage of being
time--invariant, as opposed to e.g.\ the number of observers at
any given time. All the different choices for $\fsel$ (including
MANO) suffer from an acute dependence on poorly understood
micro--physical processes involved in the evolution of life,
especially of conscious beings interested in making observations
of the fundamental constants. Our approach does not claim to
significantly improve over other treatments in this respect, but
it stands conceptually at least on equal footing. The fact that
our calculation gives an exceedingly small probability for
$\Lambda$ to be at least as large as the measured value, while
others weightings give larger probabilities, is to be seen as an
inherent failure of generic anthropic arguments.

We wish to evaluate the probability that an observer will measure
his or her universe  to have a vacuum energy density no smaller
than what we measure in our Universe. For illustrative purposes
and computability, we hold fixed all parameters of the universe
other than the vacuum energy density, considering  flat
Lemaitre-Friedmann-Robertson-Walker universes with exactly the
same matter and radiation contents and the same fluctuations as
our own at the time of matter--radiation equality. This is a
common setup in the literature, together with the assumption that
the number of observers is proportional to the baryon fraction in
halos (even though \cite{Tegmark:1997in,Tegmark:2005dy} showed
that this result is critically dependent on fixing all other
parameters). Below we show that even within this very restricted
class of models, the result is completely dependent on the
selection function one chooses. We consider only the case $\Lambda
> 0$. This restriction can only increase the probability of
observing $\Lambda$ equal to or greater than the observed one, so
we should interpret the probability we calculate as an upper
limit.

As the selection function for observing $\Lambda$ in the different
realizations we put forward the total number of observations that
observers can potentially carry out over the entire life of that
universe (called MANO for brevity, for ``Maximum Allowed Number of
Observations''). This maximum number is the product of two factors
-- the number of observers and the maximum number of observations
that each observer can make.

There is a fundamental difficulty in determining the total number
of observers, since we can neither compute nor measure it.  We
might argue, as is usually done, that it is proportional to the
number of galaxies, but the proportionality factor could be very
large, incredibly small or anywhere in between.  Indeed, if we
require, as it seems we must, that our observers be sufficiently
intelligent to make an observation of $\Lambda$, then it is not
clear  that we even know how to define our criteria, never mind
compute the probability per galaxy per unit time of meeting those
criteria. However, in the limit where observers are rare (in a way
we quantify below) the anthropic prediction for the probability of
observing $\Lambda$ will  be  independent of the density of
observers. We therefore choose instead to focus on the second
factor -- the maximum number of observations that each observer
can make.

In a $\Lambda > 0$ universe, the minimum temperature at which a
system ({\it e.g.} an observer) can operate is the de Sitter
temperature $\Tds = \rhol^{1/2}/(2 \pi \mpl)$. (Refrigerated
subsystems can run cooler, but the energy consumption of the
refrigeration more than compensates.)

As discussed in detail in \cite{Krauss:1999hj}, the maximum energy
an observer can ever collect is \be  \label{eq:Emax}E_{\rm max}
\propto \min\left(\frac{4\pi}{3}\left[(\eta_\infty - \eta_\star)
  a_\star\right]^3,n_{\rm obs}^{-1}\right)
  \rho_m(a_\star) .  \ee
Here $\eta_\star$ is the conformal time when the observer starts
collecting, while at conformal time $\eta_\infty$, $a(\eta_\infty)
= \infty$.
There are ${\cal O}(1)$ geometric prefactors that we ignore to
focus on the functional dependence of Eq.~\eqref{eq:Emax}. The
$n_{\rm obs}^{-1}$ term in \eqref{eq:Emax} represents the cutoff
in $E_{\rm max}$ due to competition with other observers. We
consider the case of ``rare observers''  -- when there is at most
one observer within the comoving volume accessible to each from
the time that they first become capable of making observations
onward. We ignore this cutoff and focus on the first term.

The number of thermodynamic processes (such as observations of
$\Lambda$) an observer can carry out is maximized if the observer
saves up $E_{\rm max} $ until the universe has reached the de
Sitter temperature. Thus \be \label{eq:nmax} \Nmax \leq E_{\rm
max}/k_B\Tds . \ee Following the arguments given above, we adopt
$\Nmax$ as a probabilistic weight in the selection function.

We start from the Friedman equation for a flat universe containing
cosmological constant, radiation and pressureless matter (both
baryonic and cold dark matter)
  \be \label{eq:fr}
  \left(\frac{a'}{a}\right)^2 = \frac{8\pi G}{3} \left(\rhol +  \rhom + \rhor\right)
  \ee
(where a prime denotes derivative with respect to  conformal time
$\eta$). We define $R L^\dagger$ to be the ratio of the vacuum
energy density in the sample universe we are considering to the
total energy density at matter--radiation equality, where
$L^\dagger\equiv(\Omega_\Lambda^\dagger)_{\rm eq}$ is the value of
this ratio in our own Universe. (Throughout, quantities with
dagger superscripts are the values measured in our Universe). We
assume that the cosmological constant is always negligible at
matter--radiation equality, i.e.\ we limit our analysis to the
regime $R \ll 10^9$. Finally, we normalize the scale factor at
matter--radiation equality, $\aeq \equiv 1$, and set $\al\equiv
a/\aeq$. It will prove  useful to note that
  \be  L^\dagger
= \frac{1}{2} \left(\Om_{\Lambda}^\dagger\right)_0 \left(1 -
\left(\Om_{\Lambda}^\dagger\right)_0\right)^{-1} \al_0^{-3}
\approx \frac{3}{2} \al_0^{-3},  \ee
  where quantities subscripted with a
$0$ are as measured in our Universe today and $\alpha_0\sim 3000$.
In the last equality we have used that
$\left(\Om_{\Lambda}^\dagger\right)_0 \approx 3/4$, which we take
henceforth as an equality.

With the above definitions, we readily compute
  \be \label{eq:timediff}
\aeq \Heq (\eta_\infty - \eta_\star)  = \int_{\al_\star}^{\infty}
\frac{ \al^{-2} d\al}{\sqrt {R L^\dagger+ \frac{1}{2}(\al^{-3} +
\al^{-4})}}.
  \ee
The matter and radiation energy densities are $\al^{-3}/2$ and
$\al^{-4}/2$ respectively  because we assumed we can ignore
$\Lambda$ at matter--radiation equality. $\Heq$ is the value of
the Hubble parameter at equality.  (So $\Heq=\Heq^\dagger$.)
Notice that having neglected $\Lambda$ at equality does not affect
our results, since the time at which observers evolve (denoted by
a subscript $\star$) is presumably much later than equality,
$\al_\star \gg 1$. We can also neglect the term proportional to
$\al^{-1}$ in \eqref{eq:timediff}, and we can then solve
analytically the Friedman equation for $a$ as a function of
physical time $t$ (as opposed to conformal time $\eta$), and so
obtain \be \label{eq:alphastar} \al_\star =
{\al_0}\left(3R\right)^{-1/3}  \sinh \left( \ln(\sqrt{3} + 2)
\sqrt{R} \tau \right)^{2/3} . \ee We have introduced $\tau \equiv
t_\star/t^\dagger_0$, the time until observers smart enough to
begin collecting energy arise, in units of 13.7 Gyrs, the age at
which such observers (us, or our descendants) are known to have
arisen in our Universe.

Before proceeding to a numerical evaluation of
\eqref{eq:timediff}, it is instructive to look at its asymptotic
limits. For the maximum number of observations,
Eq.~\eqref{eq:nmax}, we find \be \label{eq:Nmaxcrude}  \Nmax
\propto \left\{
\begin{matrix}
  \frac{2}{3}\left(\al_0/\al_\star\right)^3 R^{-2},  \quad R \gg 1,  \\
  54 R^{-1}, \quad R \ll 1.
  \end{matrix}
  \right.
  \ee
Following Eq.~\eqref{eq:prob}, we  identify $\fsel(\Lambda)
\propto \Nmax(R)$. Since we have assumed that $f(\Lambda)$ is flat
in $\Lambda$ (and hence in $R$), we have $\fobs \propto \Nmax$.
(This is the essence of our MANO weighting.) We see from
\eqref{eq:Nmaxcrude} that $\fobs$ is a steeply decreasing function
of $R$. Consequently, there is only a small anthropically
conditioned probability that $\Lambda$ is larger than we observe,
ie
  \be p(R\geq1\vert \tau)
\simeq \int_{1}^\infty \fobs(R;\tau) \ll 1. \ee
  The probability $p(R\geq1|\tau)$ can be
estimated by normalizing $\fobs$ approximately by integrating
\eqref{eq:Nmaxcrude} up to $R=1$. However, $\fobs \propto R^{-1}$
as $R \rightarrow 0$, so the normalization integral diverges
logarithmically, and is dominated by the minimum cut--off value,
$\Rmin$, if such exists. In the landscape scenario (see e.g.\
\cite{Vilenkin:2006qf}
 and references therein), for instance, the number of vacua is estimated to be of
order $10^{500}$, and therefore the corresponding minimum value of
$\Lambda$ can perhaps be taken to be $\Lambda_\text{min} \sim
10^{-500}\mpl^4$. This gives $\Rmin \sim 10^{-377}$ or
  \be
  p(R\geq1\vert \tau=1) \sim 8 \cdot 10^{-6} .
   \ee
This means that a low--$\Lambda$ universe is more probable than
one with the value of $\Lambda$ that we observe.

\begin{figure}[tb]
\centering
\includegraphics[width=0.9\linewidth]{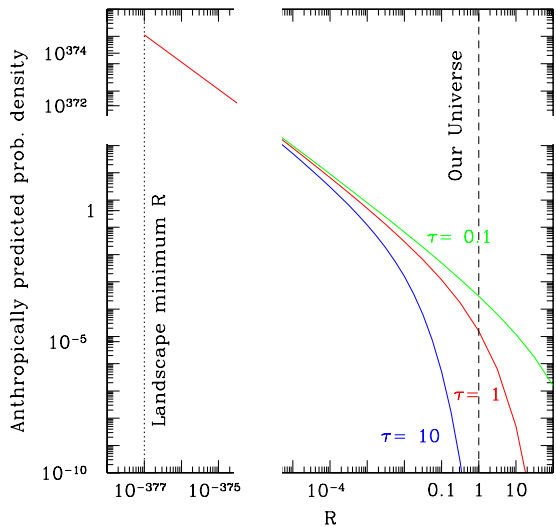}
\caption{Anthropically predicted probability density distribution
as a function of $R$, the ratio of the cosmological constant in
another part of the multiverse to the value it takes in our
Universe, in our MANO scheme and for different values of the
parameter $\tau$ controlling the cosmic time when intelligent life
emerges (in units of 13.7 Gyrs, with $\tau=1$ being the case of
our Universe). The probability of observing a value of $\Lambda$
equal to or greater than what we measure (dashed vertical line) is
very small as all the weight lies close the the minimum value. }
\label{fig:PofRvsR}
\end{figure}

In Figure \ref{fig:PofRvsR}, we plot the value of
$\fobs(R\vert\tau)$ for a few values of $\tau$ around $1$. The
probability of measuring a value of $R\geq1$ is very small, being
$9 \cdot 10^{-6}$ for $\tau = 1$ (this value is computed
numerically) and falling to $4\cdot10^{-12}$ for $\tau = 10$. The
situation is only marginally better if intelligent observers
evolve before one--tenth of the current age of the universe, since
for $\tau=0.1$ $\fobs(R\vert\tau) = 5\cdot10^{-4}$.

So far, we have worked exclusively in the rare observer limit --
where each intelligent observer is free to collect all of the
energy within their apparent horizon without competition from
other observers. One might imagine that as the density of
observers rose, one would mitigate the preference for low
$\Lambda$, but the case is by no means so clear.  If the observer
density is high, then the observers will come into competition for
the universe's (or at least their Hubble volume's) same scarce
resources.   Our own historical experience is that such
competition never leads to negotiated agreement to use those
resources as conservatively as possible.  More likely is that the
competition for resources will lead to some substantial fraction
of those resources being squandered in warfare until only one of
the observers remains.  Moreover, unless they eliminate all
possible competitors, observers will continue to spend their
finite supply of energy at a rate exceeding that which would
otherwise be necessary. What is clear is that given our inability
to predict or measure either the density of intelligent observers
or the way in which they would behave when they meet, our ability
to use anthropic reasoning can only be further compromised.

It is interesting to note that the rare observer limit would be
better described as a rare civilization limit, since the
individual observers need not be rare, but only the groups of them
that act collectively.  In this case, we could attempt to pass
from weighting by the number of possible observations to weighting
by the number of observers (although throughout all time). In the
simplest case, where the number of observations that a
civilization makes during its existence is a constant, the
translation is trivial -- the two weighting schemes are identical.
However, one can easily imagine that this number grows or
decreases with time in a way that we cannot possibly predict.
Moreover, one would still have to solve the formidable problem of
calculating the number density of civilizations.

We have argued that anthropic reasoning suffers from the problem
that the peak of the selection function depends on the details of
what exactly one chooses to condition upon -- be it the number of
observers, the fraction of baryons in halos or the total number of
observations observers can carry out. Whereas a weighting
proportional to the number density of galaxies implies that the
expected value of $\Lambda$ is close to what we observe, the
weighting scheme we propose -- according to the maximum number of
possible observations -- implies that the expected value of
$\Lambda$ is logarithmically close to its minimum allowed
non--negative value (or is zero or negative). In its usual
formulation, the anthropic principle does not offer any motivation
-- from either fundamental particle physics or probability theory
-- to prefer one weighting scheme over another. Ours is only one
specific example out of many possible weighting schemes one might
imagine (see \eg~\cite{Bousso:2006ev} for an example involving
holographic arguments). Since neither weighting scheme (nor any of
the many others one can imagine) is clearly the correct one from a
probability theory point of view (meaning one that does not lead
to paradoxical or self--contradictory conclusions of the type
described in~\cite{Neal}), we must conclude that anthropic
reasoning cannot be used to explain the value of the cosmological
constant. We expect that similar statements apply to any
conclusions that one would like to draw from anthropic reasoning.

{\em Acknowledgments} GDS is supported in part by fellowships from
the John Simon Guggenheim Memorial Foundation and Oxford's
Beecroft Institute for Particle Astrophysics and Cosmology, and by
grants from the US DoE and NASA to CWRU's particle astrophysics
theory group. RT is supported by the Royal Astronomical Society
and thanks the Galileo Galilei Institute for Theoretical Physics
for hospitality and the INFN for partial support. The authors
thank Lawrence Krauss  for extensive conversations and detailed
suggestions, in particular regarding the relation between the
number of observers and the number of observations.  They also
thank Subir Sarkar for extensive conversations.

\end{document}